\documentclass[preprint,authoryear,12pt]{elsarticle}

\usepackage{amssymb}
\usepackage{amsmath}

\journal{Computational Statistics and Data Analysis}

\usepackage{url}

\begin{document}

\begin{frontmatter}



\title{MultiDendrograms: variable-group agglomerative hierarchical clusterings}


\author[deim]{Sergio G{\'o}mez\corref{cor1}}
\ead{sergio.gomez@urv.cat}
\ead[url]{http://deim.urv.cat/~sgomez}

\author[deim]{Justo Montiel}

\author[deim]{David Torres}

\author[deq]{Alberto Fern{\'a}ndez}
\ead{alberto.fernandez@urv.cat}

\cortext[cor1]{Corresponding author. Tel.: +34 977 55 8508; fax: +34 977 55 9710.}

\address[deim]{Departament d'Enginyeria Inform{\`a}tica i Matem{\`a}tiques, Universitat Rovira i Virgili, Av.\ Pa{\"{\i}}sos Catalans 26, 43007 Tarragona, Spain}
\address[deq]{Departament d'Enginyeria Qu{\'{\i}}mica, Universitat Rovira i Virgili, Av.\ Pa{\"{\i}}sos Catalans 26, 43007 Tarragona, Spain}

\begin{abstract}
\textit{MultiDendrograms} is a Java-written application that computes agglomerative hierarchical clusterings of data. Starting from a distances (or weights) matrix, \textit{MultiDendrograms} is able to calculate its dendrograms using the most common agglomerative hierarchical clustering methods. The application implements a variable-group algorithm that solves the non-uniqueness problem found in the standard pair-group algorithm. This problem arises when two or more minimum distances between different clusters are equal during the agglomerative process, because then different output clusterings are possible depending on the criterion used to break ties between distances. \textit{MultiDendrograms} solves this problem implementing a variable-group algorithm that groups more than two clusters at the same time when ties occur.
\end{abstract}


\end{frontmatter}

% \linenumbers

\section{Introduction}
\label{sec:introduction}

Agglomerative hierarchical clustering \citep{Cormack1971, Gordon1999, Sneath1973} starts from a proximity matrix between individuals, each one forming a singleton cluster, and gathers clusters into groups of clusters or \textit{superclusters}, the process being repeated until a complete hierarchy of partitions into clusters is formed. Among the different types of agglomerative methods we find Single Linkage, Complete Linkage, Unweighted Average, Weighted Average, etc., which differ in the definition of the proximity measure between clusters. Except for the Single Linkage case, all the other agglomerative hierarchical clustering techniques suffer from a non-uniqueness problem, sometimes called the \textit{ties in proximity} problem, when the standard pair-group algorithm is used. This problem arises when two or more minimum distances between different clusters are equal during the agglomerative process. The standard approach consists in choosing any pair of clusters, breaking the ties between distances, and proceeds in the same way until a final hierarchical classification is obtained. However, different output clusterings are possible depending on the criterion used to break ties.

From the family of agglomerative hierarchical methods, Single Linkage and Complete Linkage are more susceptible than other methods to encounter ties during the clustering process, since they do not produce new proximity values different from the initial ones. With regard to the presence of ties in the original data, they are more frequent when someone works with binary variables, or even with integer variables comprising just some few distinct values. However, they can also appear using continuous variables, specially if the precision of experimental data is low. Sometimes, on the contrary, the absence of ties might be due to the representation of data with more decimal digits than it should be done. The non-uniqueness problem also depends on the measure used to obtain the proximity values from the initial variables. Moreover, in general, the larger the data set, the more ties arise \citep{MacCuish2001}.

The ties in proximity problem is well-known from several studies in different fields, for example in biology \citep{Arnau2005, Backeljau1996, Hart1983}, in psychology \citep{VanDerKloot2005}, or in chemistry \citep{MacCuish2001}. Nevertheless, this problem is frequently ignored by software packages \citep{Backeljau1996, Morgan1995, VanDerKloot2005}. Examples of packages which do not mention the problem are: the \texttt{linkage} function in the \textit{Statistics Toolbox} of MATLAB \citep{MATLAB}; the \texttt{hclust} function in the \textit{stats} package and the \texttt{agnes} function in the \textit{cluster} package of R \citep{R}; and the \texttt{cluster} and \texttt{clustermat} commands of Stata \citep{Stata}.

On the contrary, some other statistical packages warn against the existence of the non-uniqueness problem in agglomerative hierarchical clustering. For instance, if there are ties, then the \texttt{CLUSTER} procedure of SAS \citep{SAS} reports the presence of ties in the SAS log and in a column of the cluster history. The results of a cluster analysis performed with SAS depend on the order of the observations in the data set, since ties are broken as follows: ``Each cluster is identified by the smallest observation number among its members. For each pair of clusters, there is a smaller identification number and a larger identification number. If two or more pairs of clusters are tied for minimum distance between clusters, the pair that has the minimum larger identification number is merged. If there is a tie for minimum larger identification number, the pair that has the minimum smaller identification number is merged.'' \citep{SAS}.

Another example can be found in the \texttt{Hierarchical Clustering Ana\-ly\-sis} procedure of SPSS Statistics\citep{SPSS}, where it is explicitly stated that the results of the hierarchical clustering depend on the order of cases in the input file: ``If tied distances or similarities exist in the input data or occur among updated clusters during joining, the resulting cluster solution may depend on the order of cases in the file. You may want to obtain several different solutions with cases sorted in different random orders to verify the stability of a given
solution.'' \citep{SPSS}.

Finally, a third example of warning comes from the \texttt{Agglomerate} function in the \textit{Hierarchical Clustering Package} of Mathematica \citep{Mathematica}, where the user is briefly warned against the presence of ties by the following message: ``Ties have been detected; reordering input may produce a different result.'' \citep{Mathematica}.

Therefore, software packages which do not ignore the non-uniqueness problem fail to adopt a common standard with respect to ties, and they simply break ties in any arbitrary way. Here we introduce \textit{MultiDendrograms}, an application that implements the variable-group algorithm \citep{Fernandez2008} to solve the non-uniqueness problem found in the standard pair-group approach. In Section~\ref{sec:algorithms} we describe the variable-group algorithm, which groups more than two clusters at the same time when ties occur. Section~\ref{sec:manual} contains a basic manual of \textit{MultiDendrograms}. In Section~\ref{sec:soils} we show a case study performed with \textit{MultiDendrograms} using data from a real example. Finally, in Section~\ref{sec:conclusions}, we give some concluding remarks.

\section{Agglomerative hierarchical algorithms}
\label{sec:algorithms}

\subsection{Pair-group algorithm}

Agglomerative hierarchical procedures build a hierarchical classification in a bottom-up way, from a proximity matrix containing dissimilarity data between individuals of a set $\Omega = \{x_{1},\ldots,x_{n}\}$. (Note that the same analysis could be done using similarity data.) The algorithm has the following steps:
\begin{enumerate}
  \item[0)] Initialize $n$ singleton clusters with one individual in each of them: $\{x_{1}\}$, \ldots, $\{x_{n}\}$. Initialize also the distances between clusters, $D(\{x_{i}\},\{x_{j}\})$, with the values of the distances between individuals, $d(x_{i},x_{j})$:
    \begin{displaymath}
      D(\{x_{i}\},\{x_{j}\}) = d(x_{i},x_{j}) \, , \qquad \forall i,j=1,\ldots,n \, .
    \end{displaymath}
  \item[1)] Find the shortest distance separating two different clusters.
  \item[2)] Select two clusters $X_{i}$ and $X_{i'}$, subsets of $\Omega$, separated by such shortest distance and merge them into a new supercluster $X_{i} \cup X_{i'}$.
  \item[3)] Compute the distances $D(X_{i} \cup X_{i'},X_{j})$ between the new supercluster $X_{i} \cup X_{i'}$ and each of the other clusters $X_{j}$.
  \item[4)] If all individuals are not in the same cluster yet, then go back to step~1.
\end{enumerate}

Following \citet{Sneath1973}, this type of approach is known as a pair-group method, in opposition to the variable-group method which we will introduce in subsection~\ref{subsec:variable-group}. Depending on the criterion used for the calculation of distances in step~3, different agglomerative hierarchical clusterings can be implemented. The most commonly used are: Single Linkage, Complete Linkage, Unweighted Average, Weighted Average, Unweighted Centroid, Weighted Centroid, and Ward's method.

\citet{Lance1966} put these different hierarchical strategies into a single system, avoiding the need of a separate computer program for each one of them. Assume three clusters $X_{i}$, $X_{i'}$ and $X_{j}$, containing $|X_{i}|$, $|X_{i'}|$ and $|X_{j}|$ individuals respectively, and with distances between them already determined as $D(X_{i},X_{i'})$, $D(X_{i},X_{j})$ and $D(X_{i'},X_{j})$. Further assume that the smallest of all distances still to be considered is $D(X_{i},X_{i'})$, so that $X_{i}$ and $X_{i'}$ are joined to form a new supercluster $X_{i} \cup X_{i'}$, with $|X_{i}|+|X_{i'}|$ individuals. \citet{Lance1966} analyzed the distance $D(X_{i} \cup X_{i'},X_{j})$ that appears in step~3 of the above algorithm, and they expressed it in terms of the distances already defined, all known at the moment of fusion, using the following recurrence relation:
\begin{eqnarray}
  \label{eq:Lance_Williams}
  \lefteqn{ D(X_{i} \cup X_{i'},X_{j}) = \alpha_{i} D(X_{i},X_{j}) + \alpha_{i'} D(X_{i'},X_{j}) } \nonumber \\
    & & + \, \beta D(X_{i},X_{i'}) + \gamma |D(X_{i},X_{j}) - D(X_{i'},X_{j})| \, .
\end{eqnarray}
Using this technique, superclusters can always be computed from previous clusters and it is not necessary to look back at the original data matrix during the agglomerative process. The values of the parameters $\alpha_{i}$, $\alpha_{i'}$, $\beta$ and $\gamma$ determine the nature of the clustering strategy. Table~\ref{tab:Lance_Williams} gives the values of the parameters that define the most commonly used agglomerative hierarchical clustering methods.
\begin{table}
  \begin{center}
  \begin{tabular}{lccc}
    \hline
    Method              & $\alpha_{i}$ ($\alpha_{i'}$) & $\beta$ & $\gamma$ \\
    \hline
    Single Linkage      & $\frac{1}{2}$ & $0$ & $-\frac{1}{2}$ \\[1mm]
    Complete Linkage    & $\frac{1}{2}$ & $0$ & $+\frac{1}{2}$ \\[1mm]
    Unweighted Average  & $\frac{|X_{i}|}{|X_{i}|+|X_{i'}|}$ & $0$ & $0$ \\[1mm]
    Weighted Average    & $\frac{1}{2}$ & $0$ & $0$ \\[1mm]
    Unweighted Centroid & $\frac{|X_{i}|}{|X_{i}|+|X_{i'}|}$ & $-\frac{|X_{i}||X_{i'}|}{(|X_{i}|+|X_{i'}|)^{2}}$ & $0$ \\[1mm]
    Weighted Centroid   & $\frac{1}{2}$ & $-\frac{1}{4}$ & $0$ \\[1mm]
    Ward                & $\frac{|X_{i}|+|X_{j}|}{|X_{i}|+|X_{i'}|+|X_{j}|}$ & $-\frac{|X_{j}|}{|X_{i}|+|X_{i'}|+|X_{j}|}$ & $0$ \\[1mm]
    \hline
  \end{tabular}
  \end{center}
  \caption{Parameter values for Lance and Williams' formula.}
  \label{tab:Lance_Williams}
\end{table}

\subsection{Variable-group algorithm}
\label{subsec:variable-group}

The algorithm proposed by \citet{Fernandez2008} to ensure uniqueness in agglomerative hierarchical clustering has the following steps:
\begin{enumerate}
  \item[0)] Initialize $n$ singleton clusters with one individual in each of them: $\{x_{1}\}$, \ldots, $\{x_{n}\}$. Initialize also the distances between clusters, $D(\{x_{i}\},\{x_{j}\})$, with the values of the distances between individuals, $d(x_{i},x_{j})$:
    \begin{displaymath}
      D(\{x_{i}\},\{x_{j}\}) = d(x_{i},x_{j}) \, , \qquad \forall i,j=1,\ldots,n \, .
    \end{displaymath}
  \item[1)] Find the shortest distance separating two different clusters, and record it as $D_{lower}$.
  \item[2)] Select all the groups of clusters separated by shortest distance $D_{lower}$ and merge them into several new superclusters $X_{I}$. The result of this step can be some superclusters made up of just one single cluster ($|I|=1$), as well as some superclusters made up of various clusters ($|I|>1$). Notice that the latter superclusters all must satisfy the condition $D_{min}(X_{I}) = D_{lower}$, where
    \begin{displaymath}
      D_{min}(X_{I}) = \min_{i \in I} \, \min_{\substack{i' \in I \\ i' \not = i}} \, D(X_{i},X_{i'}) \, .
    \end{displaymath}
  \item[3)] Update the distances between clusters following the next substeps:
    \begin{enumerate}
      \item[3.1)] Compute the distances $D(X_{I},X_{J})$ between all superclusters, and record the minimum of them as $D_{next}$ (this will be the shortest distance $D_{lower}$ in the next iteration of the algorithm).
      \item[3.2)] For each supercluster $X_{I}$ made up of various clusters ($|I|>1$), assign a common agglomeration interval $[D_{lower},D_{max}(X_{I})]$ for all its constituent clusters $X_{i}$, $i \in I$, where
        \begin{displaymath}
          D_{max}(X_{I}) = \max_{i \in I} \, \max_{\substack{i' \in I \\ i' \not = i}} \, D(X_{i},X_{i'}) \, .
        \end{displaymath}
    \end{enumerate}
  \item[4)] If all individuals are not in the same cluster yet, then go back to step~1.
\end{enumerate}

In the same way that several agglomerative hierarchical methods can be computed with the same pair-group algorithm using Lance and Williams' formula, \citet{Fernandez2008} gave a generalization of Eq.~(\ref{eq:Lance_Williams}), compatible with the agglomeration of more than two clusters simultaneously, that can be used to compute the distances $D(X_{I},X_{J})$ in step~3 of the variable-group algorithm. Suppose we want to agglomerate two superclusters $X_{I}$ and $X_{J}$, respectively indexed by $I=\{i_{1},i_{2},\ldots,i_{p}\}$ and $J=\{j_{1},j_{2},\ldots,j_{q}\}$. Then the distance between them is defined as:
\begin{eqnarray}
  \label{eq:Fernandez_Gomez}
  \lefteqn{ D(X_{I},X_{J}) = \sum_{i \in I} \sum_{j \in J} \alpha_{ij} D(X_{i},X_{j}) } \nonumber \\
    & & + \sum_{i \in I} \sum_{\substack{i' \in I \\ i'>i}} \beta_{ii'} D(X_{i},X_{i'}) + \sum_{j \in J} \sum_{\substack{j' \in J \\ j'>j}} \beta_{jj'} D(X_{j},X_{j'}) \nonumber \\
    & & + \, \delta \sum_{i \in I} \sum_{j \in J} \gamma_{ij} [D_{max}(X_{I},X_{J}) - D(X_{i},X_{j})] \nonumber \\
    & & - (1 - \delta) \sum_{i \in I} \sum_{j \in J} \gamma_{ij} [D(X_{i},X_{j}) - D_{min}(X_{I},X_{J})] \, ,
\end{eqnarray}
where
\begin{displaymath}
  D_{max}(X_{I},X_{J}) = \max_{i \in I} \, \max_{j \in J} \, D(X_{i},X_{j})
\end{displaymath}
and
\begin{displaymath}
  D_{min}(X_{I},X_{J}) = \min_{i \in I} \, \min_{j \in J} \, D(X_{i},X_{j}) \, .
\end{displaymath}
Table~\ref{tab:Fernandez_Gomez} shows the values obtained by \citet{Fernandez2008} for the parameters $\alpha_{ij}$, $\beta_{ii'}$, $\beta_{jj'}$, $\gamma_{ij}$ and $\delta$ which determine the clustering method calculated with Eq.~(\ref{eq:Fernandez_Gomez}).
\begin{table}
  \begin{center}
  \begin{tabular}{lcccc}
    \hline
    Method              & $\alpha_{ij}$ & $\beta_{ii'}$ ($\beta_{jj'}$) & $\gamma_{ij}$ & $\delta$ \\
    \hline
    Single Linkage      & $\frac{1}{|I||J|}$ & $0$ & $\frac{1}{|I||J|}$ & $0$ \\[1mm]
    Complete Linkage    & $\frac{1}{|I||J|}$ & $0$ & $\frac{1}{|I||J|}$ & $1$ \\[1mm]
    Unweighted Average  & $\frac{|X_{i}||X_{j}|}{|X_{I}||X_{J}|}$ & $0$ & $0$ & $-$ \\[1mm]
    Weighted Average    & $\frac{1}{|I||J|}$ & $0$ & $0$ & $-$ \\[1mm]
    Unweighted Centroid & $\frac{|X_{i}||X_{j}|}{|X_{I}||X_{J}|}$ & $-\frac{|X_{i}||X_{i'}|}{|X_{I}|^{2}}$ & $0$ & $-$ \\[1mm]
    Weighted Centroid   & $\frac{1}{|I||J|}$ & $-\frac{1}{|I|^{2}}$ & $0$ & $-$ \\[1mm]
    Ward                & $\frac{|X_{i}|+|X_{j}|}{|X_{I}|+|X_{J}|}$ & $-\frac{|X_{J}|}{|X_{I}|} \frac{|X_{i}|+|X_{i'}|}{|X_{I}|+|X_{J}|}$ & $0$ & $-$ \\[1mm]
    \hline
  \end{tabular}
  \end{center}
  \caption{Parameter values for Fern{\'a}ndez and G{\'o}mez's formula.}
  \label{tab:Fernandez_Gomez}
\end{table}

Using the pair-group algorithm, only the Centroid methods (Weighted and Unweighted) may produce \textit{reversals}. Let us remember that a reversal arises in a dendrogram when it contains at least two clusters $X$ and $Y$ for which $X \subset Y$ but $h(X)>h(Y)$, where $h(X)$ is the height in the dendrogram at which the individuals of cluster $X$ are merged together \citep{Morgan1995}. In the case of the variable-group algorithm, reversals may appear in substep~3.2 when $D_{max}(X_{I}) > D_{next}$ for any supercluster $X_{I}$. Although reversals make dendrograms difficult to interpret if they occur during the last stages of the agglomerative process, it can be argued that they are not very disturbing if they occur during the first stages. Thus, as happens with the Centroid methods in the pair-group case, it could be reasonable to use the variable-group algorithm as long as no reversals at all or only unimportant ones were obtained.

In substep~3.2 of the variable-group clustering algorithm, sometimes it will not be enough to adopt a fusion interval, but it will be necessary to obtain an exact fusion value (e.g.\ in order to calculate a distortion measure). In these cases, \textit{MultiDendrograms} systematically uses the shortest distance, $D_{lower}$, as the fusion value for superclusters made up of several clusters. Such criterion allows the recovering of the pair-group result for the Single Linkage method and, in addition, it avoids the appearance of reversals. However, it must be emphasized that the use of exact fusion values, without considering the fusion intervals at their whole lengths, means that some valuable information regarding the heterogeneity of the clusters is being lost.

\section{MultiDendrograms manual}
\label{sec:manual}

This section contains a basic manual for version 3.0 of the application \textit{MultiDendrograms}. To get the latest available version of the software, please visit \textit{MultiDendrograms} web page at
\begin{center}
  \url{http://deim.urv.cat/~sgomez/multidendrograms.php}
\end{center}

\subsection{Input data}

\textit{MultiDendrograms} needs to have input data in a compatible text file format. The data file must represent a distances (or weights) matrix. There are two different arrangements that these data can be stored in such that \textit{MultiDendrograms} may accept them: matrix and list formats.

\subsubsection{Matrix-like file}

Each line in the text file contains a data matrix row. The characteristics of these files are:
\begin{itemize}
  \item The matrix must be symmetric, and the diagonal values must be zeros.
  \item Within each row, the values can be separated by: spaces (` '), tab character, semicolon (`;'), comma (`,'), or vertical bar (`$\mid$').
  \item It is possible to include labels with the names of the nodes in an additional first row or column, but not in both.
  \item If present, the labels of the nodes cannot contain any of the previous separators.
\end{itemize}

\subsubsection{List-like file}

Each line in the text file contains three fields, which represent the labels of two nodes and the distance (or weight) between them. The characteristics of these files are:
\begin{itemize}
  \item The separators between the three fields can be: spaces (` '), tab character, semicolon (`;'), comma (`,'), or vertical bar (`$\mid$').
  \item The labels of the nodes cannot contain any of the previous separators.
  \item Distances from a node to itself (e.g.\ ``\texttt{a a 0.0}'') must not be included.
  \item \textit{MultiDendrograms} accepts either the presence or absence of symmetric data lines, i.e.\ if the distance between nodes \texttt{a} and \texttt{b} is \texttt{2.0}, then it is possible to include in the list just the line ``\texttt{a b 2.0}'', or both ``\texttt{a b 2.0}'' and ``\texttt{b a 2.0}''. If both are present, the program checks whether they are equal.
\end{itemize}

\subsection{Loading data}

\begin{figure}[!t]
  \begin{center}
  \includegraphics[width=0.95\textwidth]{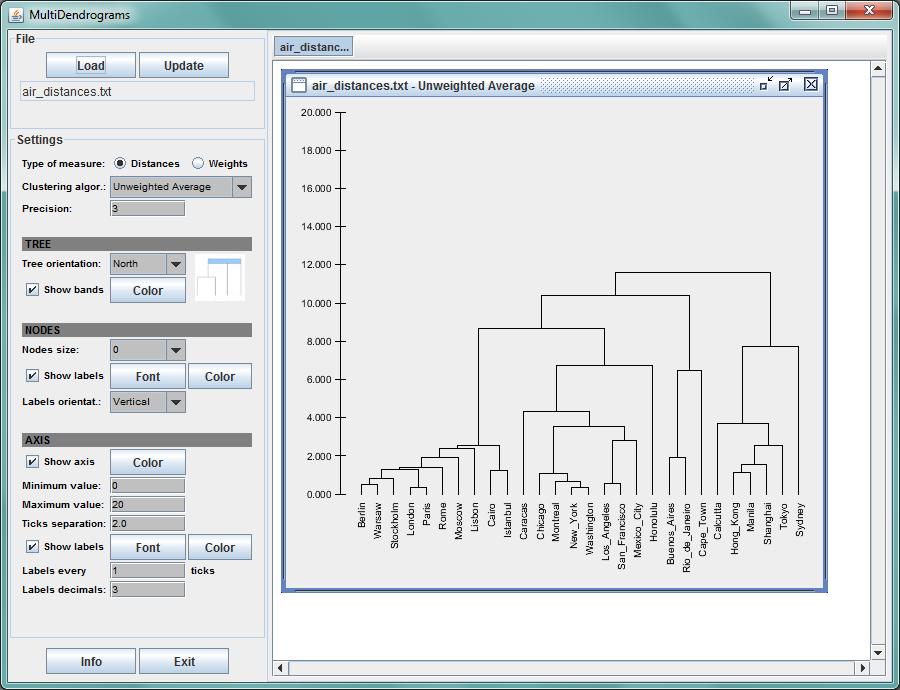}
  \end{center}
  \caption{\textit{MultiDendrograms} user interface.}
  \label{fig:load}
\end{figure}

Once we have our data in a compatible format, we can load them into \textit{MultiDendrograms}:
\begin{enumerate}
  \item Choose the desired settings, mainly the options for the \textbf{Type of measure} and the \textbf{Clustering algorithm}. These settings will be explained in detail in subsection~\ref{subsec:settings}.
  \item Click on the \textbf{Load} button.
  \item Select the file to open and then click on the \textbf{Open} button.
  \item Now the data is loaded and its multidendrogram representation is shown (see Fig.~\ref{fig:load}).
\end{enumerate}

\subsection{Actions}

\textit{MultiDendrograms} only has two action buttons: \textbf{Load} and \textbf{Update}. \textbf{Load} is used to read data from a file and create a new window for the corresponding multidendrogram, using the current values of the parameters, while \textbf{Update} is needed for the actualization of the active multidendrogram when one or more parameters are changed. Below these buttons it is shown the name of the data file corresponding to the active multidendrogram. It is possible to load the same data file several times, in order to compare the multidendrogram appearance for different parameters settings.

\subsection{Settings}
\label{subsec:settings}

The program automatically applies default values to the parameters depending on the data loaded, which should be adjusted as desired. Fig.~\ref{fig:load} shows the settings panel in the left part of the window, with four different areas corresponding to main data representation, tree, nodes and axis settings, respectively.

Changes in the main data representation parameters affect the structure of the multidendrogram tree, thus it needs to be fully recomputed, operation which may take several seconds, even minutes (depending on the data size and the computer speed). On the other hand, changes in the tree, nodes and axis settings only modify the visual representation of the multidendrogram and are much faster to update.

\subsubsection{Main data representation settings}

\begin{description}
  \item [\textbf{Type of measure}:] It allows choosing between two different types of measure, \textbf{Distances} and \textbf{Weights}. Choose between them according to the meaning of the input data. With \textbf{Distances}, the closer the nodes the lower their distance. On the contrary, with \textbf{Weights}, the closer the nodes the larger their weight. By default, \textbf{Distances} is selected.
  \item [\textbf{Clustering algorithm}:] Seven distinct algorithms are available, \textbf{Single Linkage}, \textbf{Com\-ple\-te Linkage}, \textbf{Unweighted Average}, \textbf{Weighted Average}, \textbf{Unweighted Centroid}, \textbf{Weighted Centroid}, and \textbf{Ward}. By default, \textbf{Unweighted Ave\-ra\-ge} is selected.
  \item [\textbf{Precision}:] Number of significant digits of the data to be taken into account for the calculations. This is a very important parameter, since equal distances at a certain precision may become different by increasing its value. Thus, it may be responsible of the existence of tied distances. The rule should be not to use a precision larger than the resolution given by the experimental setup that has generated the data. By default, the precision is set to that of the data value with the largest number of significant digits.
\end{description}

\subsubsection{Tree settings}

\begin{description}
  \item [\textbf{Tree orientation}:] Four orientations are available, \textbf{North}, \textbf{South}, \textbf{East}, and \textbf{West}, which refer to the relative position of the root of the tree. By default, \textbf{North} is selected.
  \item [\textbf{Show bands}:] It allows showing a band or not in case of tied minimum distances between three or more elements, and selecting the color of the band. If \textbf{Show bands} is selected, the bands show the heterogeneity of distances between the clustered elements. Otherwise, the elements are grouped at their minimum distance (see Fig.~\ref{fig:bands}). By default, \textbf{Show bands} is selected and its default color is \textbf{light gray}.
\begin{figure}[!t]
  \begin{center}
    \begin{tabular}{cc}
      \textbf{Show bands} &
      Do not \textbf{Show bands} \\
      \includegraphics[width=0.40\textwidth]{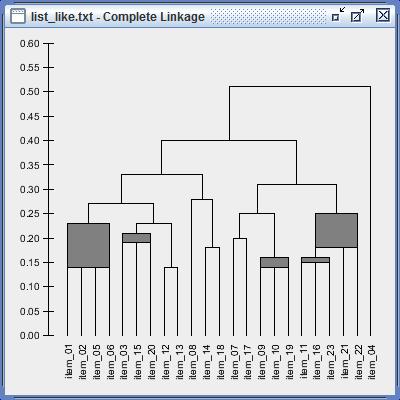} &
      \includegraphics[width=0.40\textwidth]{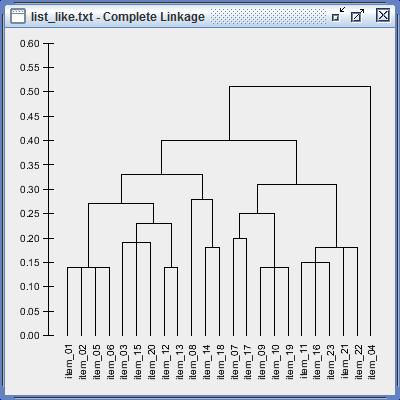}
    \end{tabular}
  \end{center}
  \caption{Possibility of showing a heterogeneity band in case of ties.}
  \label{fig:bands}
\end{figure}
\end{description}

Let us explain the meaning of the bands. In MultiDendrograms, if several pairs of elements share the same minimum distance, they are clustered together in one step. For instance, suppose that the minimum distance is 0.4 and it corresponds to the tied pairs (A, B) and (B, C). MultiDendrograms puts them together in the same cluster \{A, B, C\} at height 0.4. However, if the distance (A, C) is 0.5, it is possible to represent the cluster \{A, B, C\} as a rectangle which spans between heights 0.4 and 0.5, thus showing the heterogeneity of the clustered elements.

\subsubsection{Nodes settings}

\begin{description}
  \item [\textbf{Nodes size}:] Six different node sizes are available. By default, 0 is selected (i.e.\ nodes are not shown).
  \item [\textbf{Show labels}:] It allows showing or not the labels of the nodes and selecting their color and font. By default, \textbf{Show labels} is selected, the font is \textbf{Arial} and the color is \textbf{black}.
  \item [\textbf{Labels orientation}:] Three different orientations are available, \textbf{Vertical}, \textbf{Horizontal} and \textbf{Oblique}. By default, \textbf{Vertical} is selected.
\end{description}

\subsubsection{Axis settings}

\begin{description}
  \item [\textbf{Show axis}:] It allows showing or not the axis and selecting its color. By default, \textbf{Show axis} is selected and the color is \textbf{black}.
  \item [\textbf{Minimum value} / \textbf{Maximum value}:] They allow choosing the minimum and maximum values of the axis, respectively. They also affect the view of the multidendrogram. The default values are calculated from the loaded data.
  \item [\textbf{Ticks separation}:] It allows choosing the separation between consecutive ticks of the axis. The default value is calculated from the loaded data.
  \item [\textbf{Show labels}:] It allows showing or not the labels of the axis, and selecting their font and color. By default, \textbf{Show labels} is selected, the font is \textbf{Arial} and the color is \textbf{black}.
  \item [\textbf{Labels every \ldots ticks}:] Number of consecutive ticks to find the next labeled tick. By default is set to 1.
  \item [\textbf{Labels decimals}:] Number of decimal digits of the tick labels. By default is set equal to the \textbf{Precision} parameter.
\end{description}

\subsection{Analyzing and exporting results}

\begin{figure}[!t]
  \begin{center}
    \begin{tabular}{cc}
      \includegraphics[width=0.50\textwidth]{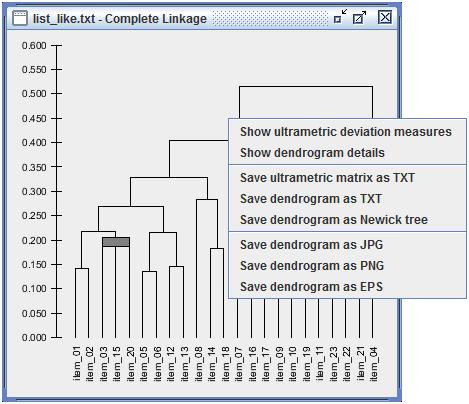}
      &
      \includegraphics[width=0.40\textwidth]{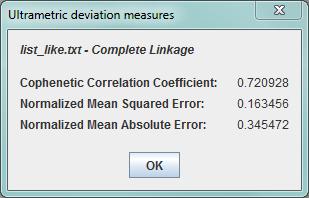}
    \end{tabular}
  \end{center}
  \caption{Contextual menu (left) and ultrametric deviation measures (right).}
  \label{fig:contextultra}
\end{figure}

The contextual menu, available by right-clicking on any multidendrogram window, gives access to several options for analyzing and exporting results to file (see Fig.~\ref{fig:contextultra}).

\begin{description}
  \item [\textbf{Show ultrametric deviation measures}:] Computes the ultrametric matrix corresponding to the active multidendrogram and obtains three deviation measures between the original matrix and the ultrametric one, which are the \textbf{Cophenetic Correlation Coeffi\-cient}, the \textbf{Normalized Mean Squared Error}, and the \textbf{Normalized Mean Absolute Error} (see Fig.~\ref{fig:contextultra}).
  \item [\textbf{Show dendrogram details}:] Opens a window that contains all the information of the multidendrogram in a navigable folder-like structure (see Fig.~\ref{fig:tree}). The available information in the details window is:
\begin{figure}[!t]
  \begin{center}
  \includegraphics[width=0.45\textwidth]{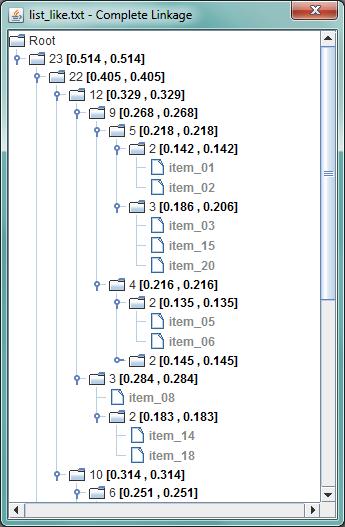}
  \end{center}
  \caption{Multidendrogram details window.}
  \label{fig:tree}
\end{figure}
  \begin{itemize}
    \item Number of data items (leafs of the tree) under each interior node of the multidendrogram. The interior nodes in the multidendrogram representation correspond to the clusters found during the agglomerative process.
    \item Minimum and maximum distances at which the children of an interior node are joined to form a new cluster. These values may only be different in case of tied minimum distances, and they become a band in the multidendrogram representation.
    \item List of children for each interior node, which may be either interior nodes or data items (leafs).
  \end{itemize}
  \item [\textbf{Save ultrametric matrix as TXT}:] Calculates the ultrametric matrix corresponding to the loaded data and saves it to a text file represented as a matrix, not as a list, with the labels of the nodes in the first row. This text file can then be easily loaded into any text editor or spreadsheet application.
  \item [\textbf{Save dendrogram as TXT}:] Saves the multidendrogram details to a text file.
  \item [\textbf{Save dendrogram as Newick tree}:] Saves the multidendrogram details in Newick tree format. In this format, the information given by the bands is lost and only the minimum distance is saved. However, it has the advantage that it is a standard format used by many other applications, thus allowing their use to generate other graphical representations.
  \item [\textbf{Save dendrogram as JPG / PNG / EPS}:] It is also possible to save the image of the multidendrogram in three different formats (JPG, PNG and EPS) using the corresponding \textbf{Save dendrogram as \ldots} context menu items.
\end{description}

\subsection{Command-line direct calculation}

It is possible to use \textit{MultiDendrograms} in command-line mode to calculate the dendrogram without the graphical interface. This is useful in several situations:
\begin{itemize}
  \item To automate the generation of many dendrograms using scripts.
  \item When there is no need of a plot of the dendrogram.
  \item When the plot of the dendrogram is to be performed with a different program.
  \item When the number of elements is too large to allow a graphical representation.
  \item To be able to call \textit{MultiDendrograms} from a different application.
\end{itemize}

The input parameters of a command-line call are:
\begin{itemize}
  \item The name of the input data file, in matrix or list format.
  \item The type of measure: distances or weights.
  \item The clustering algorithm: single linkage, complete linkage, unweighted average, weighted average, unweighted centroid, weighted centroid or ward.
  \item The precision, i.e.\ the number of significant digits of the data for the calculations. This parameter is optional, and if not given it is calculated from the data. However, the rule should be not to use a precision larger than the resolution given by the experimental setup which has generated the data.
\end{itemize}

The output results are:
\begin{itemize}
  \item A file with the dendrogram tree in text format.
  \item A file with the dendrogram in Newick format.
  \item A file with the ultrametric matrix.
  \item The ultrametric deviation measures: the cophenetic correlation coefficient, the normalized mean squared error, and the normalized mean absolute error.
\end{itemize}

The syntax of a command-line direct calculation is:
\begin{center}
  \texttt{java -jar multidendrograms.jar -direct FILE TYPE METHOD [PREC]}
\end{center}
And a concrete example for a distances matrix using the complete linkage method with 3 decimal significant digits is:
\begin{center}
  \texttt{java -jar multidendrograms.jar -direct data.txt DISTANCES Complete\_Linkage 3}
\end{center}

\section{Case study}
\label{sec:soils}

We show here a case study performed with \textit{MultiDendrograms} using data from a real example which had been previously studied by \citet{Morgan1995}, and \citet{Fernandez2008}. It is the \textit{Glamorganshire soils} example, formed by similarity data between 23 different soils. A table with the similarities between soils can be found in \citet{Morgan1995}, where the values are given with an accuracy of three decimal places. In order to work with dissimilarities as in \citet{Fernandez2008}, we have transformed the similarities $s(x_{i},x_{j})$ into the corresponding dissimilarities $d(x_{i},x_{j})=1-s(x_{i},x_{j})$, even though \textit{MultiDendrograms} is also capable of directly working with similarities.
\begin{figure}[!t]
  \begin{center}
    \begin{tabular}{l}
      (a) \\
      \includegraphics[width=0.95\textwidth]{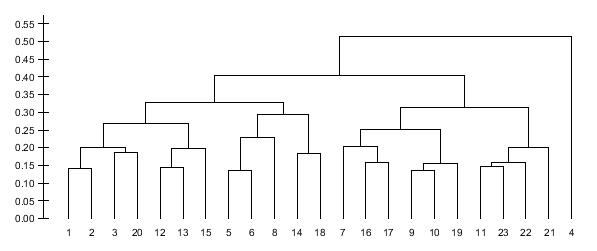} \\
      (b) \\
      \includegraphics[width=0.95\textwidth]{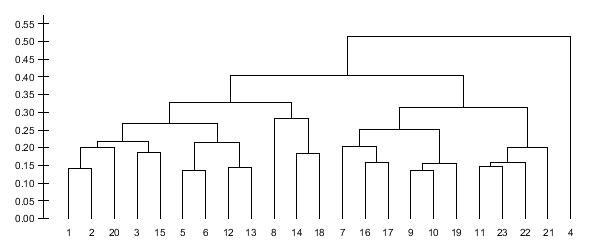}
    \end{tabular}
    \caption{Complete Linkage dendrograms for the soils data. According to the brown earths soil group formed by soils 1, 2, 6, 12 and 13, the dendrogram in~(a) is worse than the one in~(b) because the former merges these soils at a posterior stage of the agglomerative process.}
    \label{fig:soils_dendrograms}
  \end{center}
\end{figure}

The original data contain a tied value for pairs of soils (3, 15) and (3, 20), which is responsible for two different dendrograms using the Complete Linkage method (see Fig.~\ref{fig:soils_dendrograms}). \citet{Morgan1995} explain that the 23 soils had been categorized into eight ``great soil groups'' by a surveyor. Focusing on soils 1, 2, 6, 12 and 13, which are the only members of the brown earths soil group, we observe that the dendrogram in Fig.~\ref{fig:soils_dendrograms}(a) does not place them in the same cluster until they join soils from five other soil groups, forming the cluster \{1, 2, 3, 20, 12, 13, 15, 5, 6, 8, 14, 18\}. From this point of view, the dendrogram in Fig.~\ref{fig:soils_dendrograms}(b) is better, since the corresponding cluster loses soils 8, 14 and 18, each representing a different soil group. Therefore, in this case we have two possible solution dendrograms and the probability of obtaining the ``good'' one is, hence, 50\%.
\begin{figure}[!t]
  \begin{center}
    \includegraphics[width=0.95\textwidth]{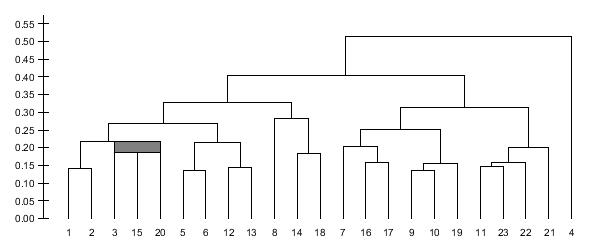}
  \end{center}
  \caption{Complete Linkage multidendrogram for the soils data.}
  \label{fig:soils_multidendrogram}
\end{figure}

We now use the \textit{MultiDendrograms} software to obtain the multidendrogram representation corresponding to the \textit{Glamorganshire soils} data. We select \textbf{Complete Linkage} as the \textbf{Clustering algorithm}, and \textbf{Precision = 3}. We obtain a multidendrogram that we have saved in JPG format using the contextual menu. The result, shown in Fig.~\ref{fig:soils_multidendrogram}, unravels the existence of a tie comprising soils 3, 15 and 20. Besides, the multidendrogram gives us the good classification, that is, the one with soils 8, 14 and 18 out of the brown earths soil group. Except for the internal structure of the cluster \{1, 2, 3, 15, 20\}, the rest of the multidendrogram hierarchy coincides with that of the dendrogram shown in Fig.~\ref{fig:soils_dendrograms}(b).

\section{Conclusions}
\label{sec:conclusions}

\textit{MultiDendrograms} is a simple yet powerful software to make hierarchical clusterings of data, distributed under an Open Source license. It implements the variable-group algorithm for agglomerative hierarchical clustering that solves the non-uniqueness problem found in the standard pair-group algorithm. This problem consists in obtaining different hierarchical classifications from a unique set of proximity data, when two or more minimum distances between different clusters are equal during the agglomerative process. In such cases, selecting a unique classification can be misleading. Software packages that do not ignore this problem fail to adopt a common standard with respect to ties, and many of them simply break ties in any arbitrary way.

Starting from a distances (or weights) matrix, \textit{MultiDendrograms} computes its dendrogram using the variable-group algorithm which groups more than two clusters at the same time when ties occur. Its main properties are:
\begin{itemize}
  \item When there are no ties, MultiDendrograms gives the same results as any pair-group algorithm.
  \item It always gives a uniquely determined solution.
  \item In the multidendrogram representation of the results, the occurrence of ties during the agglomerative process can be explicitly observed. Furthermore, the height of any fusion interval (the \textit{bands} in the program) indicates the degree of heterogeneity inside the corresponding cluster.
\end{itemize}

\textit{MultiDendrograms} also allows the tuning of many graphical representation parameters, and the results can be easily exported to file. A summary of its characteristics is:
\begin{itemize}
  \item Multiplatform: developed in Java, runs in all operating systems (e.g.\ Windows, Linux and MacOS).
  \item Graphical user interface: data selection, hierarchical clustering options, multidendrogram representation parameters, navigation across the multidendrogram, deviation measures, etc.
  \item Also command-line direct calculation without graphical user interface.
  \item Implementation of variable-group algorithms for agglomerative hierarchical clustering.
  \item Works with distance and weight matrices.
  \item Many parameters for the customization of the dendrogram layout: size, orientation, labels, axis, etc.
  \item Navigation through the dendrogram information in a folder-like window.
  \item Calculation of corresponding ultrametric matrix.
  \item Calculation of deviation measures: cophenetic correlation coefficient, normalized mean squared error, and normalized mean absolute error.
  \item Save dendrogram details in text and Newick tree format.
  \item Save dendrogram image as JPG, PNG and EPS.
\end{itemize}

Although ties need not be present in the initial proximity data, they may arise during the agglomerative process. For this reason, and given that the results of the variable-group algorithm coincide with those of the pair-group algorithm when there are no ties, we recommend to directly use \textit{MultiDendrograms}. With a single action one knows whether ties exist or not, and additionally the subsequent hierarchical classification is obtained.

\section*{Acknowledgments}

The authors thank Josep Maria Mateo for discussion and helpful comments about the practical functionalities of this software. S.G.~acknowledges support from the Spanish Ministry of Science and Innovation FIS2009-13730-C02-02, the Generalitat de Catalunya 2009-SGR-838, and the European Union FP7 FET projects PLEXMATH 317614 and MULTIPLEX 317532. A.F.~acknowledges support from the Spanish Ministry of Science and Innovation CTQ2009-14627, and the Generalitat de Catalunya 2009-SGR-1529.

\bibliographystyle{model2-names}
\bibliography{MultiDendrograms}

%% not want to use model2-names.bst.



%% \bibitem must have one of the following forms:
%%   \bibitem[Jones et al.(1990)]{key}...
%%   \bibitem[Jones et al.(1990)Jones, Baker, and Williams]{key}...
%%   \bibitem[Jones et al., 1990]{key}...
%%   \bibitem[\protect\citeauthoryear{Jones, Baker, and Williams}{Jones
%%       et al.}{1990}]{key}...
%%   \bibitem[\protect\citeauthoryear{Jones et al.}{1990}]{key}...
%%   \bibitem[\protect\astroncite{Jones et al.}{1990}]{key}...
%%   \bibitem[\protect\citename{Jones et al., }1990]{key}...
%%   \harvarditem[Jones et al.]{Jones, Baker, and Williams}{1990}{key}...
%%


% \end{thebibliography}

\end{document}